\documentstyle[twocolumn,prb,aps]{revtex}
\input epsf

\begin{document}
\draft
\twocolumn[\hsize\textwidth\columnwidth\hsize\csname@twocolumnfalse\endcsname

\title{Time-dependent, four-point density correlation function
description of dynamical heterogeneity and decoupling in supercooled
liquids}

\author{S. C. Glotzer, V.N.~Novikov\cite{iae}, and T. B. Schr\o der}
\address{Center for Theoretical and Computational Materials Science,
and Polymers Division, \\ National Institute of Standards and
Technology, Gaithersburg, MD 20899, USA}

\date{\today}
\maketitle

\begin{abstract}
Dynamical heterogeneity and the decoupling of diffusion and relaxation
in a supercooled liquid is investigated via a time-dependent,
four-point density correlation function.  We show that the main
contribution to the corresponding generalized susceptibility
$\chi_4(t)$ in a molecular dynamics simulation of a Lennard-Jones
liquid arises from spatial correlations between temporarily localized
(``caged'') particles.  By comparing $\chi_4(t)$ with a generalized
susceptibility $\chi_M(t)$ related to a correlation function for the
squared particle displacements, we demonstrate a connection
between dynamical heterogeneity and the decoupling of relaxation and
diffusion.

\end{abstract}
\vfill\eject \pacs{PACS numbers: 02.70Ns, 61.20Lc, 61.43Fs} \vskip2pc]
\narrowtext 

Spatially heterogeneous dynamics (``dynamical heterogeneity'') in
otherwise homogeneous supercooled, glassforming liquids is now well
established in experiments
\cite{sillreview,bohmreview,expts,sillescu,ediger,dino} close to the
glass transition temperature $T_g$, and this hetergeneity is even
apparent at higher temperatures above the mode coupling \cite{mct}
temperature $T_c$ in simulations \cite{sims,dgp,kdppg,dgpkp}.  For
example, recent studies \cite{dgp,kdppg,dgpkp,onuki} of the dynamics
of supercooled, glass-forming polymeric and binary simple liquids in
terms of the correlations of monomer or particle displacements
revealed the dynamical heterogeneity of these liquids and a rapidly
growing range of correlated motion on cooling towards $T_c$.  At the
same time, the decoupling of translational diffusion and relaxation as
well as translational and rotational diffusion in these fluids is also
well known, and simulations show \cite{onuki,kob,benn,schilling} that
this decoupling begins well above $T_c$ where dynamical heterogeneity
first appears.  Several authors have argued that the decoupling of
diffusion and relaxation is a direct result of dynamical
heterogeneity, with the slowest particles dominating structural
relaxation and the fastest particles dominating diffusion
\cite{sillreview,sillescu,ediger,dgpkp,onuki,stillinger,douglas}.

In this Letter we use a four-point time correlation function of the
density to probe dynamical heterogeneity in a glass-forming liquid,
and elucidate the connection between this heterogeneity and the
decoupling of bulk transport processes.  This four-point function was
first investigated in a supercooled liquid by Dasgupta, et
al. \cite{dasgupta91}, and recently Donati, et al. \cite{dfpg} have
demonstrated analytically and computationally the interesting behavior
of the related generalized four-point susceptibility $\chi_4(t)$
(defined below).  As shown in Ref.~\cite{dfpg}, $\chi_4(t)$ can be
represented in terms of the fluctuations of an "order parameter" that
is a bilinear, time-dependent product of densities. Here we show that
the self-part of $\chi_4(t)$ is directly related to spatial
correlations between temporarily localized particles, while the
distinct part is related to the correlated motion of particles into
positions previously occupied by neighboring particles.  We evaluate
these quantities for a cold Lennard-Jones (LJ) liquid, and show that
in this system $\chi_4(t)$ is dominated by growing spatial
correlations between localized particles.  We then compare the
behavior of $\chi_4(t)$ with a generalized time-dependent
susceptibility related to a correlation function of squared particle
displacements. From these two quantities we find two {\it different}
characteristic time scales: the time scale on which
temporarily localized particles are most spatially correlated scales
with temperature like the structural relaxation time, while the
time scale on which the correlation between squared
particle displacements is strongest scales like the inverse diffusion
coefficient.  In this way, we demonstrate that the decoupling of
diffusion and relaxation in this model liquid arises from dynamical
heterogeneity.

Consider a liquid of $N$ particles in a volume $V$, with density $\rho
({\bf r},t)=\sum_{i=1}^{N} \delta ({\bf r}-{\bf r}_{i}(t))$.  The
simplest density correlation function that contains information on
correlated particle motion is fourth order. We write this function in
terms of the deviations of $\rho({\bf r},t)$ from its average value, $\Delta
\rho ({\bf r},t)=\rho ({\bf r},t)-\rho _{0},$ where $\rho _{0}=\langle
\rho \rangle =N/V$, and $\langle \ldots \rangle $ denotes an ensemble
average:
\begin{eqnarray}
{\cal F}_{4}({\bf r}_{1},{\bf r}_{2},t)=\langle \Delta \rho ({\bf r}_{1},0)\Delta
\rho ({\bf r}_{1},t)\Delta \rho ({\bf r}_{2},0)\Delta \rho ({\bf r}
_{2},t)\rangle \nonumber \\
-\big\langle \Delta \rho ({\bf r}_{1},0)\Delta \rho ({\bf r}_{1},t) \big\rangle
\big\langle \Delta \rho ({\bf r}_{2},0)\Delta \rho ({\bf r}_{2},t)\big\rangle .
\label{eq1}
\end{eqnarray}
Terms involving one position only are subtracted in Eq.~(\ref{eq1})
since they contain no information on spatial correlations of
particle motions.  ${\cal F}_{4}({\bf r}_{1},{\bf r}_{2},t)$ can 
be written,
\begin{eqnarray}
{\cal F}_{4}({\bf r}_{1},{\bf r}_{2},t)={\cal G}_{4}({\bf r}_{1},{\bf r}_{2},t) + \Delta
{\cal F}_{4}({\bf r}_{1},{\bf r}_{2},t), \nonumber
\end{eqnarray}
where the two-point, two-time, fourth-order
correlation function of densities ${\cal G}_4$ is defined as
\cite{dasgupta91,dfpg}
\begin{eqnarray}
{\cal G}_{4}({\bf r}_{1},{\bf r}_{2},t)\equiv &&\langle \rho ({\bf r}_{1},0)\rho
({\bf r} _{1},t)\rho ({\bf r}_{2},0)\rho ({\bf r}_{2},t)\rangle \nonumber \\
&&-\langle \rho ({\bf r}_{1},0)\rho ({\bf r}_{1},t)\rangle \langle \rho ({\bf
r }_{2},0)\rho ({\bf r}_{2},t)\rangle. \nonumber
\end{eqnarray}
$\Delta {\cal F}_4({\bf r}_1, {\bf r}_2, t)$ consists of second- and
third-order correlation functions of density. A straightforward
calculation shows that $\int \int d{\bf r}_{1}d{\bf r}_{2}\Delta {\cal
F}_4({\bf r}_1, {\bf r}_2, t)$ vanishes by symmetry, and as a result,
the volume integrals of ${\cal F}_4({\bf r}_1, {\bf r}_2, t)$ and
${\cal G}_4({\bf r}_1, {\bf r}_2, t)$ are equal to each other and
correspond to the {\it same} generalized susceptibility $\chi^0_4(t)$,
\begin{eqnarray}
\chi^0_4 (t) = 
{\beta V \over N^2}\int \int d{\bf r}_{1}d{\bf r}_{2}{\cal G}_{4}({\bf r}_{1},{\bf
r}_{2},t).
\nonumber
\end{eqnarray}
It is straightforward to show that $\chi^0_4(t)$ can be written as 
\begin{equation}
\chi^0_4 (t)= {\beta V \over N^2} \big[ {\langle Q_{0}^{2}(t)\rangle -\langle
Q_{0}(t)\rangle ^{2}} \big],
\label{eq5}
\end{equation}
where $\beta = 1/k_BT$, and the time-dependent ``order parameter''
$Q_0(t)$ equals
\begin{eqnarray}
Q_{0}(t) =\int d{\bf r}\rho ({\bf r},0)\rho ({\bf r},t)
=\sum_{i=1}^N \sum_{j=1}^N
\delta ({\bf r}_{i}(0)-{\bf r}_{j}(t)). \label{eq6}
\end{eqnarray}

In a simulation, $Q_0(t)$ is numerically ill-defined (for a finite
system) since the probability that particle $j$ exactly replaces
particle $i$ is infinitely small.  Following Parisi \cite{parisi}, we
therefore modify $Q_0(t)$ by an ``overlap'' function $w(r)$ that is unity inside
a region of size $a$ and zero otherwise, where $a$ is taken on the
order of a particle diameter \cite{dfpg,cardenas}. This leads to an
$a$-dependent counterpart to $Q_0(t)$,
\begin{eqnarray}
Q(t) &=&\int d{\bf r}_{1}d{\bf r}_{2}\rho ({\bf r}_{1},0)\rho ({\bf r}
_{2},t)w(| {\bf r}_{1}-{\bf r}_{2}|). \nonumber \\ 
&=&\sum_{i=1}^N \sum_{j=1}^N
\int d{\bf r}w(|{\bf r}|)\delta ({\bf r}+{\bf r}_{i}(0)-
{\bf r}_{j}(t))  \nonumber \\
&=&\sum_{i=1}^N \sum_{j=1}^N
w(|{\bf r}_{ij}-{\vec{\mu}}_{j}|)  \label{eq8} 
\end{eqnarray}
where ${\bf r}_{ij}\equiv{\bf r}_{i}(0)-{\bf r}_{j}(0)$ and
${\vec{\mu}_{i}\equiv{\bf r}_{i}(t)-{\bf r}_{i}(0)}$ is the
displacement of particle $i$ during the time interval from zero to
$t$. We choose $a=0.3\sigma_{AA}$ as in Ref.~\cite{dfpg}.

Replacing $Q_{0}(t)$ in Eq.~(\ref{eq5}) by $Q(t)$ yields 
\begin{equation}
\chi_4 (t)= {\beta V \over N^2} \big[ {\langle Q^{2}(t)\rangle -\langle
Q(t)\rangle ^{2}} \big],
\label{eqchi}
\end{equation}
which gives the following expression \cite{dfpg} for $\chi_4(t)$ in
terms of the four-point correlation function $G_{4}({\bf r}_{1},{\bf
r}_{2},{\bf r}_{3},{\bf r}_{4},t)$:
\begin{eqnarray}
\chi_4 (t)= {\beta V \over N^2} \int &d&{\bf r}_{1}d{\bf r}_{2}d{\bf
r}_{3}d{\bf r}_{4}w(|{\bf r} _{1}- {\bf r}_{2}|)w(|{\bf r}_{3}-{\bf
r}_{4}|) \nonumber \\ 
&& \times \ \ G_{4}({\bf r}_{1},{\bf r}_{2},{\bf r}_{3},{\bf r}_{4},t)
\label{eq14a}
\end{eqnarray}
where
\begin{eqnarray}
G_{4}({\bf r}_{1},{\bf r}_{2},{\bf r}_{3},{\bf r}_{4},t) = && \langle
\rho ({\bf r}_{1},0)\rho ({\bf r}_{2},t)\rho ({\bf r}_{3},0)\rho (
{\bf r}_{4},t)\rangle \nonumber \\ && -\langle \rho ({\bf
r}_{1},0)\rho ({\bf r}_{2},t)\rangle \langle \rho ({\bf r }_{3},0)\rho
({\bf r}_{4},t)\rangle .
\label{eq14d}
\end{eqnarray}

We can write $Q$ in terms of its self and distinct parts, $
Q=Q_{S}+Q_{D}.$ The self part $Q_{S}$ corresponds to terms with $i=j$ in
Eq.~(\ref{eq8}):
\begin{eqnarray}
Q_{S}(t)  = \sum_{i}^N\int d{\bf r}w(r{\bf )}\delta ({\bf r}+{\bf r}
_{i}(0)-{\bf r}_{i}(t)) = \sum_{i}^N w(\mu _{j}),  \label{eq12} 
\end{eqnarray}
where $\mu_j$ is the magnitude of ${\vec{\mu}}_j$.  The distinct part
$Q_{D}$ is equal to
\begin{equation}
Q_{D}(t)= \sum_{i=1}^N \sum_{{j=1}\atop{j\neq i}}^N
w(|{\bf r}_{ij}-\vec{\mu}_{j}|). \nonumber
\end{equation}
Then $\chi_4 (t)$ can be decomposed into self ($\chi_{SS}$), distinct
($\chi_{DD}$), and interference ($\chi_{SD}$) parts: $\chi
=\chi_{SS}+\chi_{DD}+\chi_{SD}$.  From Eq.~(5), $\chi_{SS}$ and
$\chi_{DD}$ describe the fluctuations of $Q_{S}$ and $ Q_{D}$,
respectively, and $\chi_{SD}$ describes the cross fluctuations:
$\chi_{SS}\propto \langle Q_{S}^{2}\rangle -\langle Q_{S}\rangle ^{2}
$, $\chi_{DD}\propto \langle Q_{D}^{2}\rangle -\langle Q_{D}\rangle
^{2}$, and $\chi_{SD}\propto \langle Q_{S}Q_{D}\rangle -\langle
Q_{S}\rangle \langle Q_{D}\rangle$.  According to Eq.~(\ref{eq12}),
$Q_{S}(t)$ contains only contributions from small displacements, $\mu
_{i}<a$, since $w(\mu_i)=0$ for $\mu_i>a$, and thus $\chi_{SS}(t)$ is
the susceptibility of localized particles, those which during a time
interval $[0,t]$ move less than a distance $a$.  In contrast,
$Q_{D}(t)$ contains contributions from particles for which $|{\bf
r}_{ij}-{ \vec{\mu}}_{j}| < a$; that is, particles that replace
neighboring particles in a time interval $[0,t]$. 

In Ref.~\cite{dgp}, a different generalized susceptibility
$\chi_U(t)$ was defined in terms of the fluctuations in an ``order
parameter'' given by the total particle displacement in a time
interval $t$, $U(t)= \sum_{i=1}^N \mu_i(t) =\int d{\bf r} \, u({\bf
r},t)$, where the displacement density field $u({\bf
r},t)=\sum_{i=1}^N\mu _{i}(t)\delta ({\bf r}-{\bf r}_{i}(0)).$ In a
similar fashion, Eq.~(\ref{eq12}) can be rewritten as $Q_{S}(t)=\int
d{\bf r} \, q_{S}({\bf r},t)$ where the ``localization density'' field
$q_S({\bf r},t)$ is $ q_{S}({\bf r},t)=\sum_{i=1}^Nw(\mu _{i}(t){\bf
)}\delta ({\bf r}-{\bf r} _{i}(0)).$ Here we compare $\chi_4(t)$ with
$\chi_M(t)$, defined as
\begin{eqnarray}
\chi_M(t) = {{\beta V}\over \langle M(t) \rangle^2} \big[ \langle
M^{2}(t)\rangle -\langle M(t)\rangle ^{2} \big]
\label{eq21}
\end{eqnarray}
where $M(t) \equiv \sum_{i=1}^N\mu^2_{i}(t)$ (i.e. $M(t)$ is the sum
of the {\it squared} displacements for one system in a time interval
[0,t]). Like $\chi_U(t)$, $\chi_M(t)$ is proportional to the volume
integral of a correlation function of (in this case squared) particle
displacements \cite{dgp,lancaster}.  Note that both the displacement
density field $u({\bf r},t)$ and squared-displacement density field
$m({\bf r},t)$ are dominated by delocalized, or mobile particles,
while the localization density field $q_{S}({\bf r},t)$ is dominated
by localized, or immobile particles.

To evaluate these quantities we use data obtained from a molecular
dynamics simulation of a model LJ glass-former. The system is a
three-dimensional binary mixture (50:50) of 500 particles interacting
via LJ interaction parameters \cite{simul}. We analyze data from state
points at seven different temperatures $T$ approaching $T_c \approx
0.592$ from above \cite{t2} at a constant density $\rho \approx
1.3$. (In the remainder of this paper, all values are quoted in
reduced units \cite{simul}.)  All quantities presented here are
evaluated in the $NVE$ ensemble following equilibration of the system
at each state point. Further simulation details may be found in
\cite{t,t2}.

\begin{figure}[tbp]
\hbox to\hsize{\epsfxsize=1.0\hsize\hfil\epsfbox{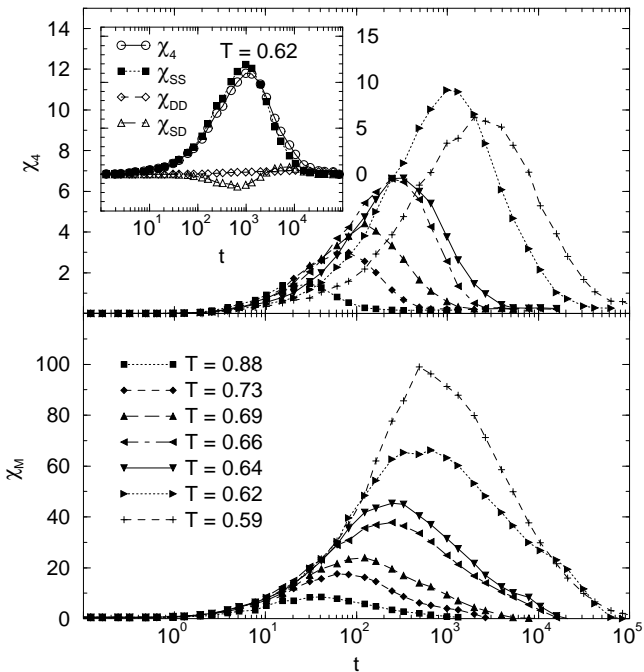}\hfil}
\caption{ (a) Time dependence of the susceptibility $\chi_4(t)$ at
various temperatures, as indicated in (b).  Inset: Self, distinct and
cross terms of $\chi_4(t)$ at $T=0.62$. (b) Time dependence of the
``squared-displacement'' susceptibility $\chi_M(t)$ at
the same values of $T$ as in (a).}
\label{fig1}
\end{figure}

In Fig.~\ref{fig1}a the susceptibility $\chi_4(t)$ calculated via
Eq.~(\ref{eqchi}) is shown as a function of time for different values
of $T$.  As found for a different LJ mixture in Ref.~\cite{dfpg}, for
all $T$, $\chi_4(t)$ is zero at short time and attains a small
constant value at large time, and has a maximum at some intermediate
time $t_{4}^{*}$.  Both $t_{4}^{*}$ and the amplitude of the peak,
$\chi_4(t^*_4)$, increase strongly with decreasing $T$.  At the lowest
value of $T$, the amplitude of $\chi_4(t)$ decreases, possibly due to
finite size effects or to the change in dynamics \cite{t2} near $T_c$.
The inset shows the self, distinct and cross terms of $\chi_4(t)$ for
one value of $T$, and we see that $\chi_{SS}$ is indeed the dominant
term. Thus, $\chi_4(t)$ is dominated by the growing range of spatial
correlations between {\it localized} particles in this fluid, and
$t_4^*$ is the time when this correlation is strongest.  In fact,
several authors have reported evidence of a growing length scale
associated with solid-like behavior in dense fluids \cite{solid}.

Fig.~1b shows $\chi_M(t)$ calculated from Eq.~\ref{eq21}, as a
function of time for different values of $T$. We find that $\chi_M(t)$
becomes negligable at small and large times and has a maximum at some
intermediate time $t_{M}^{*}$ where the spatial correlation of squared
particle displacements is strongest. This behavior is similar to that
exhibited by $\chi_U(t)$ calculated in Ref.~\cite{dgp}.

As shown in Fig.~2, both $\chi_4(t_4^*)$ and $\chi_M(t_M^*)$ increase
strongly with decreasing $T$ (with the exception of $\chi_4(t_4^*)$ at
the lowest temperature).  Over the limited temperature range of our
simulations, both functions may be reasonably fitted by power-law
functions $(T-T_{c})^{-\gamma}$ with $T_c =0.592$, with the apparent
exponents $\gamma_4 = 0.80 \pm 0.07$ and $\gamma_M = 0.87 \pm 0.05$,
as shown in the figure. (In fitting the power law, $T_c$ is held fixed
to the value $T_c=0.592$ determined in previous work \cite{t2}.)  Of
course, precise determination of the functional form requires
simulations at lower temperatures, and larger simulations to reduce
any possible finite size effects expected due to the growing range of
correlated particle motion and localization driving the growth
of $\chi_M(t)$ and $\chi_4(t)$, respectively.

\begin{figure}[tbp]
\hbox to\hsize{\epsfxsize=1.0\hsize\hfil\epsfbox{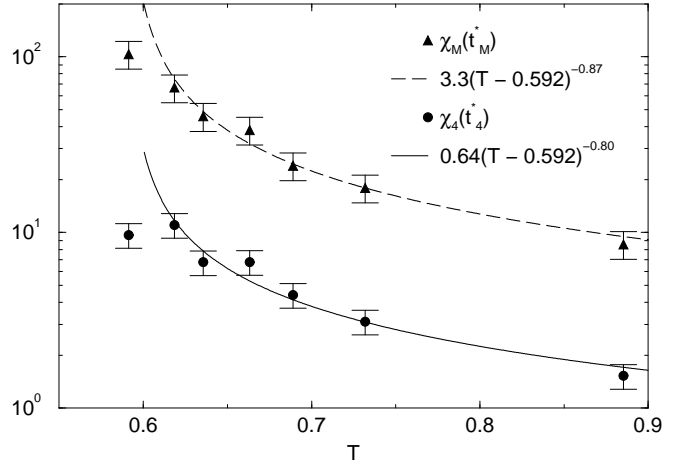}\hfil}
\caption{Temperature dependence of $\chi_4(t_4^*)$ and
$\chi_M(t_M^*)$.  The solid and dashed lines are power law fits to the
data as indicated (excluding the lowest temperature). The error bars
are estimated from deviations between three independent samples, where
for each sample, $\chi_4(t)$ and $\chi_M(t)$ are calculated by averaging over
128 independent time origins.}
\label{fig2}
\end{figure}

Figs.~3(a) and (b) show the $T$-dependence of $t_4^*$ and $t_M^*$,
respectively, and compare them with both the inverse self-diffusion
coefficient $D^{-1}$ and the structural relaxation time
$\tau_{\alpha}$. (Here D is calculated from the mean square
displacement for the $B$ (small) particles, and $\tau_{\alpha}$ is
calculated by fitting the $\alpha$-relaxation part of the self
intermediate scattering function for the $A$ (large) particles (not
shown, see Ref.~\cite{t2}) by a stretched exponential function.)  Also
shown are power law fits to $D^{-1}$ and $\tau_{\alpha}$ excluding the
lowest temperature (see Ref.~\cite{t2} for details).  Diffusion and
relaxation are found to be ``decoupled'' in this cold liquid, as
observed in many other real and simulated cold liquids
\cite{sillescu,ediger,onuki,kob,adg}. In the present system, we find
that with $T_c=0.592$, $\gamma_D = 1.11 \pm 0.03$ and $\gamma_{\tau}=
1.41 \pm 0.07$.  Remarkably, we find \cite{also} that the
$T$-dependence of $t_{4}^{*}$ coincides within our numerical error
with that of $\tau_\alpha$, while $t_M^*$ behaves like $D^{-1}$.  That
is, the time scale on which the localized particles are most spatially
correlated scales with temperature like the structural relaxation
time, and the time scale on which the correlation between squared
particle displacements is strongest scales like the inverse diffusion
coefficient.  Thus, our data demonstrates that the ``decoupling'' of
diffusion and relaxation (or viscosity) may be directly attributed to
the emergence of dynamical heterogeneity, as argued by, e.g., Sillescu
and coworkers \cite{sillescu}, Ediger and coworkers \cite{ediger},
Stillinger \cite{stillinger},  and Douglas \cite{douglas}.

\begin{figure}[tbp]
\hbox to\hsize{\epsfxsize=1.0\hsize\hfil\epsfbox{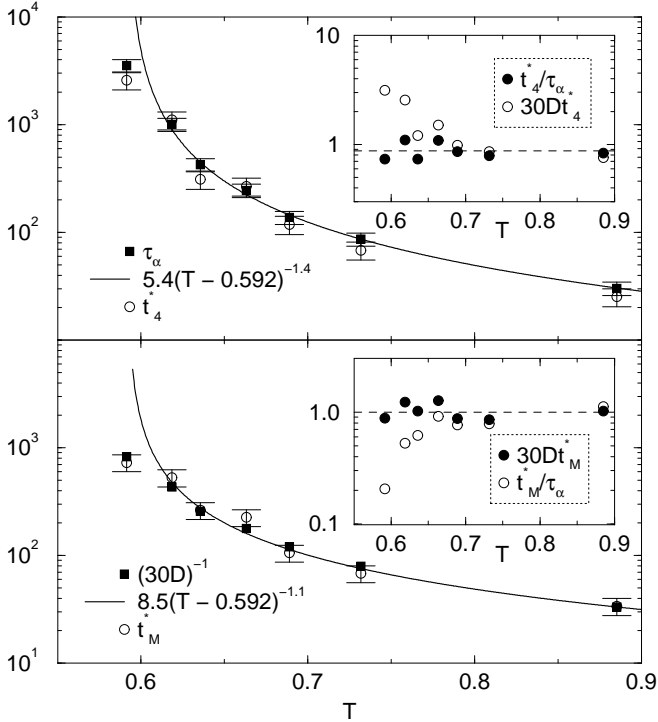}\hfil}
\caption{(a) Temperature dependence of $\tau_{\alpha}$ and the time
$t_4^*$ at which $\chi_4(t)$ exhibits a maximum.  (b) Temperature
dependence of the inverse self-diffusion coefficient D and the time
$t^*_M$ at which $\chi_M(t)$ exhibits a maximum.  The solid lines are
power law fits to $\tau_{\alpha}$ and $D^{-1}$ respectively (excluding
the lowest temperature), with $T_c$ fixed.  Insets: Comparison of
$t_4^*$ and $t_M^*$ with both $D^{-1}$ and $\tau_{\alpha}$. As
plotted, a line of zero slope (dashed line) indicates proportionality.}
\label{fig3}
\end{figure}

Our results demonstrate the importance of time-dependent higher order
density correlation functions in the characterization of dynamical
heterogeneity in supercooled liquids, and the ramifications of this
heterogeneity for the bulk dynamics.  In particular, the increasing
amplitude of the generalized time-dependent susceptibility $\chi_4(t)$
with decreasing $T$, as shown also in Ref.~\cite{dfpg}, demonstrates
an essential difference between two- and four-point density
correlation functions in these fluids.  For a glass-forming LJ liquid,
we have shown that $\chi_4(t)$ is dominated by growing spatial
correlations between temporarily localized particles.  Finally, we
have demonstrated that the decoupling of diffusion and structural
relaxation observed in supercooled liquids follows naturally from
dynamical heterogeneity, as discussed by Sillescu, Ediger, 
Stillinger and Douglas: the time scale for spatial correlations of localized
particles to develop governs structural relaxation, while the
(different) time scale for the development of spatial correlations of
squared particle displacements, which is dominated by mobile
particles, governs diffusion.  We note that it should be possible to
determine the 4-point functions studied here in colloidal suspensions using
particle tracking methods.

We thank C. Donati, J.F. Douglas, S. Franz, R.D. Mountain, G. Parisi,
P.H. Poole, and F. Starr for valuable comments on the manuscript.


\begin{references}
\bibitem[{*}]{iae}  On leave from Institute of Automation \& Electrometry,
Russian Academy of Sciences, Novosibirsk, 630090, Russia.

\bibitem{sillreview} H. Sillescu, J. Non-Cryst. Solids {\bf 243}, 81 (1999), and 
references therein.

\bibitem{bohmreview} R. B\"ohmer, Curr. Opin. Sol. State
Mater. Sci. {\bf 3}, 378 (1998) and references therein.

\bibitem{expts} See, e.g., K. Schmidt-Rohr and H.W. Spiess,
Phys. Rev. Lett. {\bf 66}, 3020 (1991); A. Heuer, et al.,
Phys. Rev. Lett. {\bf 95}, 2851 (1995); R. Richert,
J. Non-Cryst. Solids, {\bf 172}-{\bf 174}, 209 (1994); R. B\"ohmer, et
al., J. Non-Cryst. Solids {\bf 235-237}, 1 (1998); F. R. Blackburn, et
al., J. Non-Cryst. Solids {\bf 172-174}, 256 (1994).
  
\bibitem{sillescu} F. Fujara, B. Geil, H. Sillescu, and G. Fleischer,
Z. Phys. B {\bf 88}, 195 (1992); I. Chang, et al.,
J. Non-Cryst. Solids {\bf 172-174}, 248 (1994); H. Sillescu,
Phys. Rev. E {\bf 53}, 2992 (1996); I. Chang and H. Sillescu,
J. Phys. Chem. B {\bf 101}, 8794 (1997).

\bibitem{ediger} M.T. Cicerone, F.R. Blackburn and M.D. Ediger,
Macromolecules {\bf 28}, 8224 (1995); M.T. Cicerone and M.D. Ediger,
J. Chem. Phys. {\bf 104}, 7210 (1996).

\bibitem{dino} L. Andreozzi, A. Di Schino, M. Giordano, and
D. Leporini, Europhys. Lett. {\bf 38}, 669 (1997).

\bibitem{sims} See, e.g., T. Muranaka and Y. Hiwatari, Phys.Rev.E {\bf
51}, R2735 (1995); R.D. Mountain, J.Chem.Phys. {\bf 102}, 5408,
(1995); M. Hurley and P. Harrowell, Phys. Rev. E {\bf 52}, 1694
(1995); D. Perera and P. Harrowell.  J. Non-Cryst. Sol. {\bf 235-237},
314 (1998); B. Doliwa and A. Heuer, Phys. Rev. Lett. {\bf 80}, 4915
(1998); A. Onuki and Y. Yamamoto, J. Non-Cryst. Sol. {\bf 235-237}, 34
(1998).

\bibitem{dgp}  C. Donati, P.H. Poole, and S.C. Glotzer, Phys. Rev. Lett.,
{\bf 82}, 6064 (1999); C. Bennemann, C. Donati, J. Baschnagel and S. C. Glotzer,
Nature {\bf 399}, 246 (1999).

\bibitem{kdppg} S. C. Glotzer and C. Donati, J. Phys.: Cond. Matt. 11, A285
(1999); W. Kob, C. Donati, P.H. Poole, S.J. Plimpton and S.C. Glotzer, 
Phys. Rev. Lett. {\bf 79}, 2827 (1997). 

\bibitem{dgpkp}  C. Donati, S. Glotzer, P. Poole, W. Kob, and S. Plimpton,
Phys. Rev. E {\bf 60}, 3107 (1999).

\bibitem{mct}  W. G\"{o}tze, L. Sj\"{o}gren, Rep. Progr. Phys. {\bf 55}, 241
(1992).

\bibitem{onuki} R. Yamamoto and A. Onuki, Phys. Rev. Lett. {\bf 81},
4915 (1998); Phys. Rev. E {\bf 58}, 3515 (1998).

\bibitem{kob} W. Kob and H.C. Andersen, Phys. Rev. E {\bf 51}, 4626 (1995).

\bibitem{benn}  C. Bennemann, W. Paul, K. Binder, and B. D\"{u}nweg, Phys.
Rev. E {\bf 57}, 843 (1998).

\bibitem{schilling} S. Kammerer, W. Kob and R. Schilling, Phys. Rev. E {\bf 56}, 5450 (1997).

\bibitem{stillinger} F. H. Stillinger and J. A. Hodgdon, Phys. Rev. E
{\bf 50}, 2064 (1994); Phys. Rev. E {\bf 53}, 2995 (1996). 

\bibitem{douglas} J.F. Douglas and D. Leporini, J. Non-Crystalline Sol. {\bf 235-237}, 137 (1998).

\bibitem{dasgupta91} C. Dasgupta, A.V. Indrani, S. Ramaswamy and
M.K. Phani, Europhys. Lett. 15 307 (1991).

\bibitem{dfpg} C. Donati, S. Franz, G. Parisi, and S.C. Glotzer, 
cond-mat/9905433.

\bibitem{parisi} G. Parisi, J. Phys. A: Math. Gen. {\bf 30},
L765-L770 (1997). 

\bibitem{cardenas} M. Cardenas, S. Franz, and G. Parisi, J. Phys. A:
Math. Gen. {\bf 31}, L163 (1998); J. Chem. Phys. {\bf 110}, 1726
(1999).

\bibitem{lancaster} D. Lancaster and G. Parisi, J. Phys. A:
Math. Gen. {\bf 30}, 5911 (1997).

\bibitem{simul} The LJ interaction parameters are $\sigma _{BB}/\sigma
_{AA}=5/6$, $\sigma _{AB}=(\sigma _{AA}+\sigma _{BB})/2$, and $
\epsilon _{AA}=\epsilon _{AB}=\epsilon _{BB}$. The masses are given by
$ m_{B}/m_{A}=1/2$. The length of the sample is $L=7.28\sigma _{AA}$
and the potential was cut and shifted at $2.5\sigma _{\alpha \beta
}$. All quantities are reported in reduced units: $T$ in units of
$\epsilon _{AA}$, lengths in units of $\sigma _{AA}$ and time in units
of $\tau \equiv (m_{B}\sigma _{AA}^{2}/48\epsilon )^{1/2}$. Adopting
Argon parameters for the A particles gives $\tau =3\times 10^{-13}\
$s. 

\bibitem{t}  T.B. Schr\o der and J.C. Dyre, J. Non-Cryst. Solids {\bf 235-237},
331 (1998).

\bibitem{t2}  T.B. Schr\o der, S. Sastry, J.C. Dyre and S.C. Glotzer,
cond-mat/9901271.

\bibitem{solid} R.D. Mountain, ACS Symposium Series {\bf
676}, 122 (1997); R. Ahluwalia and S.P. Das, Phys. Rev. E {\bf 57},
5771 (1998); G. Johnson, et al., Phys. Rev. E {\bf 57}, 5707 (1998).

\bibitem{adg} P. Allegrini, J.F. Douglas and S.C. Glotzer,
Phys. Rev.~E, in press (cond-mat/9907313).

\bibitem{also} We also calculated $\chi_U(t)$ following
Refs.~\cite{dgp}; we find that although $t^*_u$ scales more like
$D^{-1}$ than like $\tau_{\alpha}$, $t^*_M$ scales more like $D^{-1}$
than does $t^*_u$. 

\end{references}
\end{document}